\documentclass[lettersize,journal]{IEEEtran}
\usepackage{amsmath,amsfonts}
\usepackage{algorithmic}
\usepackage{algorithm}
\usepackage{array}
\usepackage[caption=false,font=normalsize,labelfont=sf,textfont=sf]{subfig}
\usepackage{textcomp}
\usepackage{stfloats}
\usepackage{url}
\usepackage{verbatim}
\usepackage{graphicx}
\usepackage{cite}
\usepackage{booktabs}
\usepackage{multirow}
\usepackage{makecell}
\usepackage{caption}
\usepackage{enumitem}
\usepackage{xcolor}
\usepackage{amssymb}
\usepackage{pifont} 
\usepackage[pagebackref=true,breaklinks=true,colorlinks,bookmarks=false]{hyperref}

\newcommand{\cmark}{\textcolor{green}{\ding{51}}}
\newcommand{\xmark}{\textcolor{red}{\ding{55}}}

\begin{document}

\title{Audio-Plane: Audio Factorization Plane Gaussian Splatting for Real-Time Talking Head Synthesis}

\author{Shuai Shen, Wanhua Li, Yunpeng Zhang, Yap-Peng Tan,~\IEEEmembership{Fellow,~IEEE}, and Jiwen Lu,~\IEEEmembership{Fellow,~IEEE}
\thanks{Shuai Shen and Yap-Peng
Tan are with the School of Electrical and Electronic Engineering, Nanyang Technological University, Singapore, 639798. (Email: shuai.shen@ntu.edu.sg; eyptan@ntu.edu.sg).}
\thanks{Wanhua Li is with the John A. Paulson School of Engineering and Applied Sciences at Harvard University, Cambridge, MA, USA. (Email: wanhua016@gmail.com).}
\thanks{Yunpeng Zhang is with the PhiGent Robotics, Beijing, China. (Email: yunpengzhang97@gmail.com).}
\thanks{Jiwen Lu is with the Beijing National Research Center for Information Science and Technology (BNRist) and the Department of Automation, Tsinghua University, Beijing, 100084, China. (Email: lujiwen@tsinghua.edu.cn). (Corresponding author: Jiwen Lu)}
}



\maketitle

\begin{abstract}
Talking head synthesis has emerged as a prominent research topic in computer graphics and multimedia, yet most existing methods often struggle to strike a balance between generation quality and computational efficiency, particularly under real-time constraints.
In this paper, we propose a novel framework that integrates Gaussian Splatting with a structured Audio Factorization Plane (Audio-Plane) to enable high-quality, audio-synchronized, and real-time talking head generation. For modeling a dynamic talking head, a 4D volume representation, which consists of three axes in 3D space and one temporal axis aligned with audio progression, is typically required. However, directly storing and processing a dense 4D grid is impractical due to the high memory and computation cost, and lack of scalability for longer durations.
We address this challenge by decomposing the 4D volume representation into a set of audio-independent spatial planes and audio-dependent planes, forming a compact and interpretable representation for talking head modeling that we refer to as the Audio-Plane. 
This factorized design allows for efficient and fine-grained audio-aware spatial encoding, and significantly enhances the model's ability to capture complex lip dynamics driven by speech signals.
To further improve region-specific motion modeling, we introduce an audio-guided saliency splatting mechanism based on region-aware modulation, which adaptively emphasizes highly dynamic regions such as the mouth area. This allows the model to focus its learning capacity on where it matters most for accurate speech-driven animation. Extensive experiments on both the self-driven and the cross-driven settings demonstrate that our method achieves state-of-the-art visual quality, precise audio-lip synchronization, and real-time performance, outperforming prior approaches across both 2D- and 3D-based paradigms. We highly recommend viewing our demonstration video at \url{https://sstzal.github.io/Audio-Plane/} for intuitive visual comparisons and qualitative results.
\end{abstract}


\section{Introduction}

\IEEEPARstart{A}{udio}-driven talking head synthesis aims to generate realistic talking face videos that precisely synchronize lip movements with the input speech, while preserving the speaker’s identity and visual consistency. This technology has rapidly evolved into a significant research focus in computer graphics, vision, and multimedia systems~\cite{zakharov2019few,zhang2020davd,wang2021one,chen2020talking,zhou2020makelttalk,sargin2008analysis}, due to its potential to enable lifelike digital avatars and natural human-computer interaction. Its wide range of applications spans virtual reality, online education, telepresence, digital entertainment, and assistive communication, making it a key enabler for the next generation of immersive and personalized media experiences.

\begin{figure*}[t]
  \centering
   \includegraphics[width=1\linewidth]{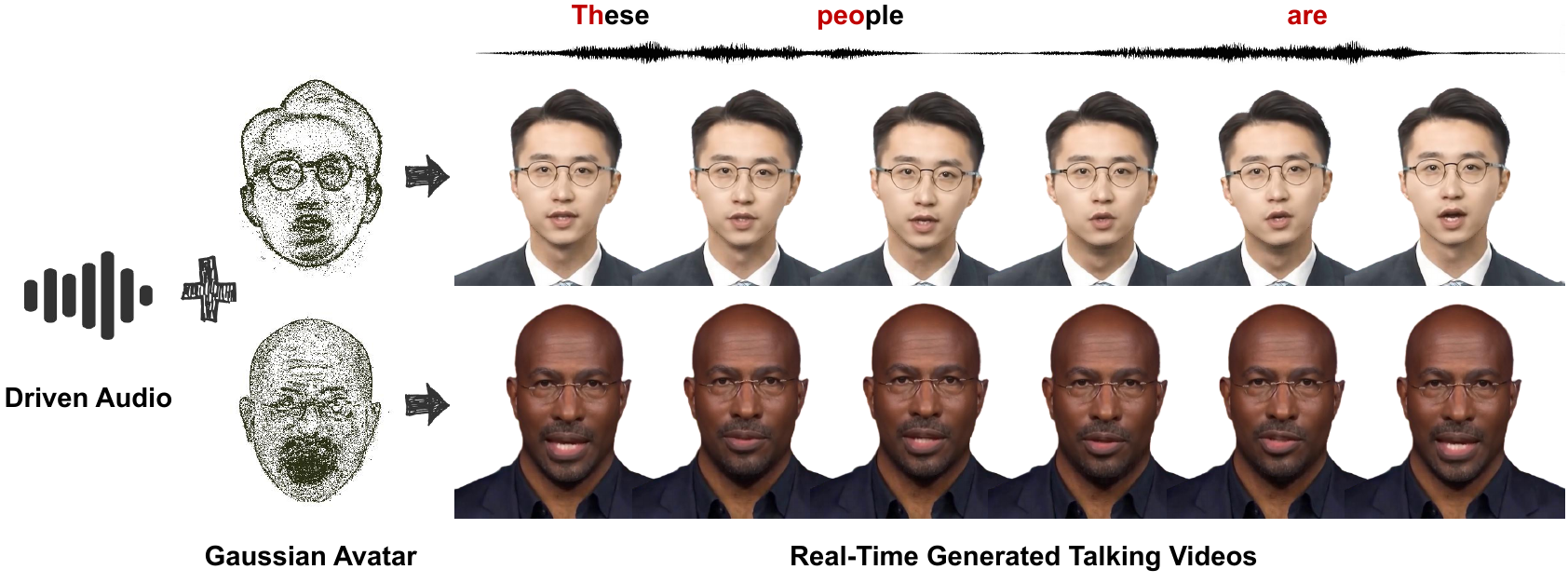}
\vspace{-4mm}
\caption{We present a compact and interpretable Audio-Plane representation into the Gaussian Splatting framework for high-quality and efficient talking head synthesis. By driving head Gaussians with audio, our method can synthesize vivid and temporally synchronized talking videos in real time.}
\label{fig1_1}
\vspace{-2mm}
\end{figure*}

In recent years, the field of talking head synthesis has seen significant advancements, with growing focus on enhancing generation quality and audio-lip synchronization. These existing works can be broadly categorized into 2D-based talking head synthesis methods and 3D-based ones. Predominantly, 2D-based approaches utilize Generative Adversarial Networks (GANs) or face warping techniques~\cite{das2020speech,gu2020flnet,christos2020headgan,prajwal2020lip,lu2021live,li2023one} to map audio signals to corresponding lip movements. While these models can generate plausible results, their lack of explicit 3D facial geometry often leads to artifacts and limits their ability to produce high-resolution and temporally consistent outputs. More recently, the introduction of Diffusion Model-based talking head synthesis methods have notably improved the generation fidelity, however these models still struggle with the control of 3D facial structures.
In contrast, 3D-based methods~\cite{zollhofer2018state,thies2016face2face,guo2021ad,kwak2022injecting,zhu2021detailed} explicitly model facial geometry and head motion in three-dimensional space, leading to more realistic and temporally stable talking head videos.
Notable technical frameworks in this category include 3D Morphable Models (3DMM)~\cite{blanz1999morphable} and Neural Radiance Fields (NeRFs)~\cite{mildenhall2020nerf}, both of which have significantly advanced the photorealistic talking head generation. Nevertheless, the practical deployment of these models remains challenging due to their high computational cost and slow inference speed.

Balancing generation quality with computational efficiency has consistently been a critical challenge in the development of talking head synthesis systems. More recently, the emergence of Gaussian Splatting~\cite{kerbl20233d,wu20234d} has opened up new possibilities for real-time and high-fidelity 3D-aware rendering, offering a promising alternative to traditional volumetric and neural rendering techniques such as 3DMM~\cite{blanz1999morphable} and NeRFs~\cite{mildenhall2020nerf}. In this work, we explore the potential of this emerging paradigm by introducing an Audio Factorization Plane-based Gaussian Splatting framework for audio-driven talking head synthesis. Our method leverages the expressiveness and rendering efficiency of Gaussian primitives while integrating rich audio-conditioned dynamics into the generation process.

\begin{figure*}[t]
  \centering
   \includegraphics[width=1\linewidth]{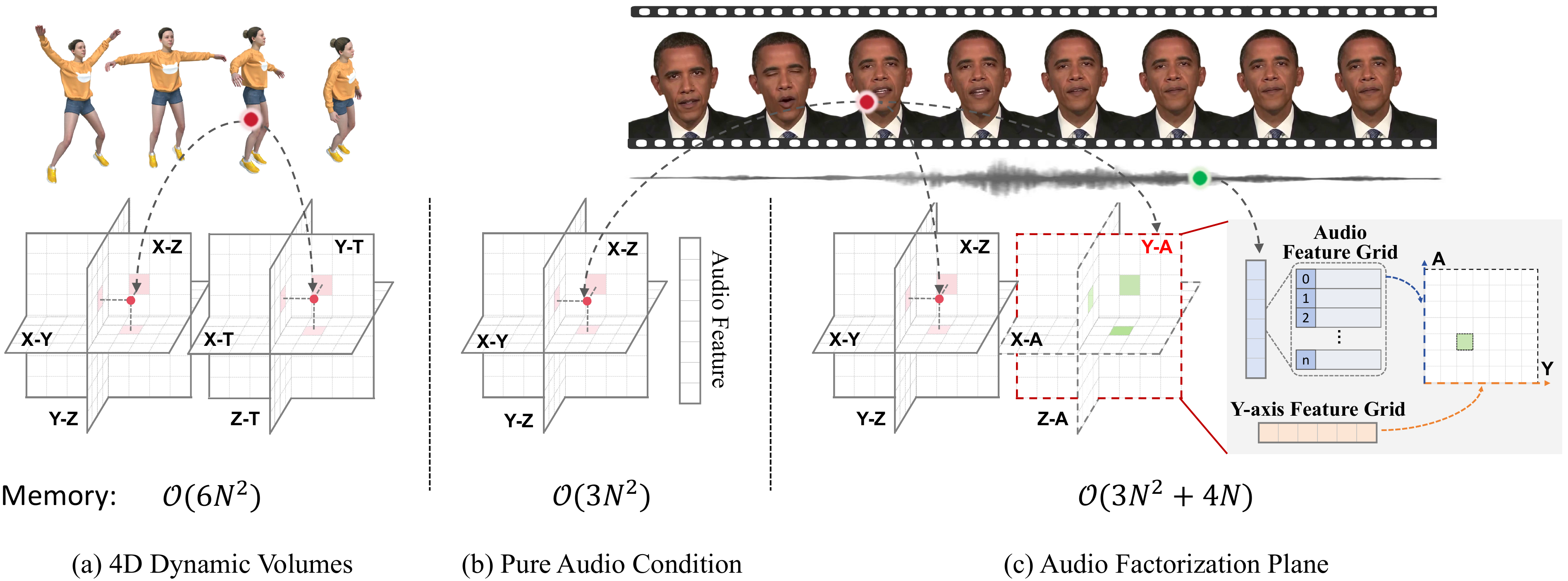}
\vspace{-2mm}
\caption{Illustration and memory usage comparison of the proposed Audio Factorization Plane and two alternative representations for dynamic scene modeling. Our method introduces a more compact (\emph{i.e.}, low memory usage) and interpretable representation for audio-driven dynamics, enabling fine-grained, spatially-aware modulation conditioned on audio. This design also maintains the audio-spatial alignment in 3D space, contributing to improved synthesis fidelity. A comprehensive qualitative comparison from four perspectives, including \emph{low memory usage}, \emph{interpretability}, \emph{audio-driven synthesis}, and \emph{audio-spatial alignment} of the three modeling paradigms is provided in Table~\ref{tab_overview}.}
\label{fig1}
\vspace{-2mm}
\end{figure*}

To faithfully model the dynamic talking head, a 4D volume representation that spans 3D spatial space and time (where the temporal time axis is implicitly governed by the input audio signal in this task) is essential. However, maintaining a dense 4D grid poses significant computational and memory challenges, particularly when scaling to long-duration sequences or high-resolution outputs. These constraints underscore the need for a compact, interpretable, and scalable 4D representation that can efficiently encode dynamic facial motions with less redundancy and better scalability.
Fig.~\ref{fig1} illustrates some 4D scene representations and their memory usage. (a) is the conventional six-planes methods~\cite{cao2023hexplane,fridovich2023k} that extend tri-planes~\cite{chan2022efficient,chen2022tensorf} with time-conditioned planes to represent the dynamic scene. Such methods rely on explicit temporal indexing to model dynamics, typically assuming that motion evolves along a continuous scalar time axis. However, in audio-driven talking head synthesis, the input is a high-dimensional and temporally structured audio signal, which cannot be represented as a simple scalar or continuous index. Consequently, this fundamental mismatch prevents the straightforward application of temporally-indexed six-planes methods to audio-conditioned facial animation tasks.
Fig.~\ref{fig1} (b) shows the common baseline~\cite{tang2022real, li2023efficient,chan2022efficient} for talking head synthesis, where audio and tri-plane spatial features are learned independently, and then concatenated as input to the subsequent network.
This shallow late fusion strategy struggles to model fine-grained and location-specific audio-visual interactions. Moreover, the global audio representation lacks spatial awareness, limiting its ability to modulate localized geometry. This design also reduces interpretability and places a greater burden on downstream networks to learn complex cross-modal relationships implicitly.
Building on the insights of previous dynamic scene representations~\cite{cao2023hexplane,fridovich2023k,li2023efficient}, we introduce the Audio Factorization Plane (Audio-Plane) as Fig.~\ref{fig1} (c).
The proposed Audio-Plane representation decomposes the audio-aware 4D volume into two components: three audio-independent spatial planes and three audio-conditioned modulation planes.
A core challenge in constructing audio-conditioned planes lies in the nature of audio features, which are high-dimensional vectors rather than scalar quantities, making it non-trivial to directly incorporate them as coordinate axes. To address this, we further factorize each audio-conditioned plane into a pair of interacting grids: an audio prototype grid and a spatial feature grid. Each pair spans an implicit feature plane, such as the illustrative \textit{Y–A} plane shown in (c). Here we use dashed lines to depict \textit{Y–A}, emphasizing that it is not an explicit feature plane but rather spanned by two constituent grids.
An overview of these three types of dynamic scene modeling methods is shown in Table~\ref{tab_overview}. We compare them across four perspectives, including \emph{low memory usage} (compact), \emph{interpretability}, \emph{audio-driven synthesis}, and \emph{audio-spatial alignment} (`A-S Align'). In comparison, our factorized design offers a compact, interpretable representation of audio-driven dynamics with superior audio-spatial alignment, enabling fine-grained, audio-aware spatial modulation.
Furthermore, to enhance the dynamic modeling of mouth regions, we introduce an audio-guided saliency splatting method to explicitly supervise the dynamism of 3D Gaussian points. Specifically, we assign each Gaussian point a learnable dynamism weight to construct the deformation field. These weights are then projected into the image space to enable explicit region-aware deformation supervision, guiding the model to emphasize motion in highly dynamic areas while suppressing unnecessary deformation in static regions.

\begin{table}[t]
    \centering
    \caption{Overview comparison of the three dynamic scene modeling paradigms listed in Fig.~\ref{fig1}. `A-S Align.' refers to Audio-Spatial Alignment. In comparison, our method achieves low memory usage, strong interpretability, effective audio-driven synthesis, and superior audio-spatial alignment. This demonstrates the overall effectiveness and robustness of our approach across multiple key dimensions.}
    \renewcommand\tabcolsep{5pt}
    \begin{tabular}{c c c c c c}
    \toprule
         &Low Memory&Interpretability&Audio-Driven&A-S Align.\\
         \midrule
         (a)&\xmark&\cmark&\xmark&\xmark\\
         (b)&\cmark&\xmark&\cmark&\xmark\\
         Ours&\cmark&\cmark&\cmark&\cmark\\
    \bottomrule
    \end{tabular}
    \vspace{-2mm}
    \label{tab_overview}
\end{table}

By integrating the Audio Factorization Plane representation with the proposed region-aware saliency splitting mechanism into the Gaussian Splatting framework, our method can achieve high-fidelity and real-time talking video generation with precise audio-driven facial motion control as shown in Fig.~\ref{fig1_1}.
Comprehensive experiments show the superiority of the proposed method in rendering quality and speed, which provides a strong baseline for talking head synthesis. To summarize, we make the following contributions:

\begin{itemize}
  \item We propose the \textbf{Audio Factorization Plane} representation, which integrates temporal audio features with spatial grids through a structured, plane-based decomposition. This 
  explicitly capture the spatially-aware influence of speech, and enables compact and interpretable representation for dynamic facial modeling.
  \vspace{1.1mm}
  \item We introduce an \textbf{Audio-Guided Saliency Splatting} strategy that explicitly supervises the motion behavior of 3D Gaussian points via region-aware deformation guidance. This design 
  emphasize motion in expressive regions while suppressing deformation in static areas, facilitating more precise control over localized facial dynamics.
  \vspace{1.1mm}
  \item Extensive experiments demonstrate that by integrating the Audio Factorization Plane and Audio-Guided Saliency Splatting mechanism into the Gaussian Splatting framework, our method can generate vivid talking videos in real time, establishing a strong baseline for high-quality and efficient audio-driven facial animation.
\end{itemize}

\section{Related Work}
In this section, we briefly review recent developments in some related topics, including 2D-based and 3D-based audio-driven talking-head synthesis methods and the emerging Gaussian Splatting technology.

\subsection{2D-Based Audio-driven Talking Head Synthesis}
The task of audio-driven talking head synthesis aims to generate realistic talking videos synchronized with the input audio. Existing approaches can be typically categorized into 2D-based and 3D-based ones, depending on their underlying representation and modeling strategy. The 2D-based methods~\cite{ prajwal2020lip,das2020speech,gu2020flnet,chen2019hierarchical,doukas2021headgan,wu2018reenactgan,vougioukas2020realistic,tang2024memories} usually treat this task as a process of audio-to-lip translation, and primarily employs some technical frameworks including landmark-driven warping~\cite{poux2021dynamic,zhou2020makelttalk,guo2024liveportrait}, GAN-based generation~\cite{hou2021mw,wu2021f3a,christos2020headgan,yu2022cmos,hong2023dagan,doukas2023free}, and emerging diffusion models~\cite{shen2023difftalk,wang2024headdiff,tian2024emo,xu2024hallo,cui2024hallo2}.
Zhang~\textit{et al.}~\cite{chen2019hierarchical} propose a hierarchical cross-modal framework for talking face generation, which introduces a dynamic pixel-wise loss to enhance lip synchronization and visual fidelity across temporal scales.
Chung~\textit{et al.}~\cite{chung2017you} develop an encoder-decoder CNN model to generate talking faces with a joint embedding of the face and audio. 
Vougioukas~\textit{et al.}~\cite{vougioukas2020realistic} present an end-to-end model with temporal GANs for more natural and coherent videos.
Wav2lip~\cite{prajwal2020lip} proposes an encoder-decoder CNN model with a well-trained lip-sync expert to generate lip-syncing talking faces.
MakeItTalk~\cite{zhou2020makelttalk} disentangles the facial information and audio content to enable speaker-aware talking-head animations. Shen~\textit{et al.}~\cite{shen2023difftalk} integrate audio-visual conditions into a diffusion-based generative framework, achieving high-resolution talking video synthesis with strong generalization across diverse identities.
However, since 2D-based methods do not incorporate explicit 3D geometry, they face several inherent limitations. First, the lack of geometric constraints may lead to unstable training and temporal inconsistencies, especially under large head poses or complex expressions. Moreover, these methods typically struggle to maintain facial structure fidelity and identity consistency, particularly in high-resolution generation.

\subsection{3D-Based Audio-driven Talking Head Synthesis}
Compared to 2D-based audio-driven talking head synthesis methods, 3D-based approaches usually achieve higher fidelity and improved temporal consistency in facial animation. In earlier studies, a series of works~\cite{suwajanakorn2017synthesizing,linsen2020ebt,thies2020neural,wang2024styletalk,wang2011reconstructing} utilized classic 3D Morphable Models~\cite{blanz1999morphable} as the intermediate representations. Suwajanakorn~\textit{et al.}~\cite{suwajanakorn2017synthesizing} leverage a large dataset of real Obama speeches to train a neural network that learns to map audio features to mouth shapes based on a 3D face model. 
Thies~\textit{et al.}~\cite{thies2020neural} propose Neural Voice Puppetry, which maps input audio to 3D facial expression parameters via a dedicated Audio2Expression network, enabling realistic and temporally stable facial reenactment.
Recently, Neural Radiance Fields (NeRFs)~\cite{mildenhall2020nerf} have been employed as a new popular 3D-aware implicit representation~\cite{guo2021ad,chan2021pi,yao2022dfa,shen2022dfrf,li2023efficient,tang2022real,sheng2023toward,zhou2024animatable}. 
AD-NeRF~\cite{guo2021ad} proposes the first audio-driven neural radiance field framework for talking-head synthesis. 
DFRF~\cite{shen2022dfrf} conditions the radiance field on the facial image for few-shot synthesis. 
RAD-NeRF~\cite{tang2022real} develops a decomposed audio-spatial encoding module for talking face modeling. ER-NeRF~\cite{li2023efficient} introduces a tri-plane hash representation coupled with a region attention module for more efficient rendering. While prior works~\cite{tang2022real,li2023efficient} incorporate spatially-aware audio features, they typically apply audio conditioning at a coarse or global level. In contrast, our method factorizes audio into a phoneme prototype feature grid and embeds it along spatial dimensions, enabling fine-grained and structured audio-spatial interactions.
Although NeRF-based methods can produce high-quality visual results, their rendering speed remains far from meeting real-time requirements. More recently, Gaussian Splatting has emerged as a more computationally efficient alternative while maintaining competitive visual fidelity. This makes it a promising foundation for real-time audio-driven talking head generation.

\subsection{Gaussian Splatting}
Splatting has been an important technique in computer graphics. One of its early applications was in the field of point-based graphics, where SurfaceSplatting~\cite{zwicker2001surface} demonstrates its utility for surface rendering from point samples, emphasizing its capability to produce high-quality images without the need for complex meshing processes. 
Recent advancements in NeRFs~\cite{mildenhall2020nerf} have inspired the exploration of Gaussian Splatting (GS) techniques in the domain of differentiable volumetric rendering\cite{niemeyer2020differentiable}, aiming to enhance the efficiency and quality of 3D scene representation.
A significant milestone in this direction is 3D Gaussian Splatting (3DGS)~\cite{kerbl20233d}, which shows that GS can be effectively leveraged to render high-quality radiance fields in real-time~\cite{ren2024octree,shen2025gamba}.
Subsequent works~\cite{luiten2023dynamic,xu2023omniavatar,yang2023real,ling2024align,li2024spacetime} have extended 3DGS to the modeling of dynamic scenes, pushing its capabilities beyond static environments. 
Among them, the representative method 4DGS~\cite{wu20234d} proposes a HexPlane-based~\cite{cao2023hexplane} Gaussian deformation field, enabling accurate and efficient modeling of temporally varying 3D geometry.
Wang~\textit{et al.}~\cite{ling2024align} introduce a text-to-4D generation framework, proving the ability of Gaussian Splatting for efficient and temporally coherent dynamic scene synthesis.
Given its rendering efficiency and high visual quality, Gaussian Splatting has also been explored in the context of talking head synthesis~\cite{yu2024gaussiantalker,cho2024gaussiantalker,li2024talkinggaussian}. Cho~\textit{et al.}~\cite{cho2024gaussiantalker} develop a spatial-audio attention module to facilitate the dynamic deformation of facial regions. Li~\textit{et al.}~\cite{li2024talkinggaussian} model the face and inside mouth branches separately for more efficient training.
In this work, we focus on constructing a more compact and interpretable feature representation for audio-driven Gaussian Splatting, aiming to improve efficiency and audio-visual alignment in talking head generation.

\begin{figure*}[t]
  \centering
   \includegraphics[width=1\linewidth]{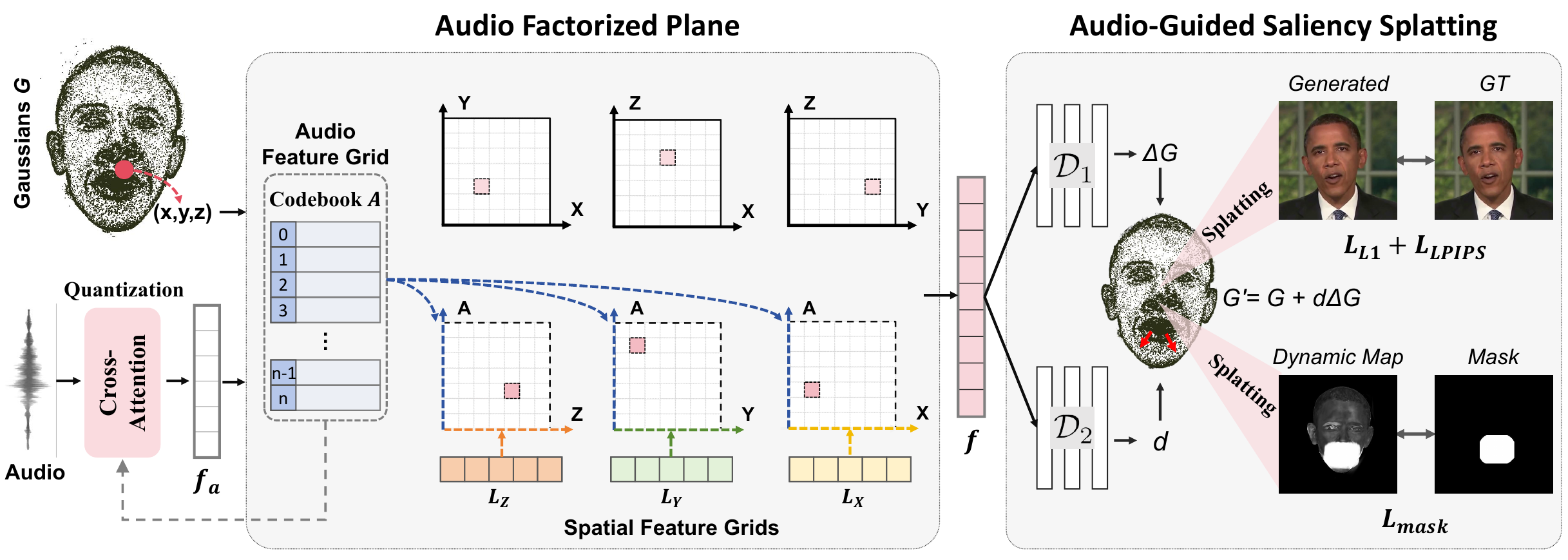}
\vspace{-2mm}
\caption{Overview of the proposed dynamic audio-driven Gaussian splatting framework for real-time talking head video synthesis. We develop the Audio Factorized Plane to incorporate audio signals into the spatial feature planes, to strengthen the model for temporally coherent facial motion modeling. We further develop an Audio-Guided Saliency Splatting method to explicitly supervise the dynamism of 3D Gaussian points, which helps to enhance the modeling of dynamic facial regions while reducing redundant complexity in less deformable areas.}
\label{fig2}
\end{figure*}

\section{Methodology}

\subsection{Overview}
\label{overview}
In this work, we propose to model the talking head as dynamic audio-driven Gaussians with an Audio Factorization Plane (Audio-Plane), and employ the differentiable splitting operation for high-quality and real-time talking video rending. An overview of the proposed method is shown in Fig.~\ref{fig2}. We present the Audio Factorization Plane by
decomposing the audio-aware 4D volume representation into a series of audio-independent spatial planes and audio-conditioned modulation planes. The proposed Audio-Plane provides a compact and interpretable representation for talking head synthesis, strengthening audio-spatial alignment and improving the model's ability to generate temporally consistent facial motion. 
To further enhance the modeling of mouth regions for better audio-lip synchronization, we introduce an audio-guided saliency splatting technique that enables explicit region-aware supervision on the motion dynamics of unordered Gaussian points. In this way, our model can better focus on the highly dynamic areas while suppressing unnecessary deformation in static regions.
In the following, a preliminary description of Gaussian Splatting is provided in Section~\ref{pre}.
We introduce the proposed dynamic Gaussian Splatting framework for talking head synthesis in Section~\ref{DGS}. In Section~\ref{Sptial}, we explain the designed Audio Factorization Plane. Section~\ref{weight} details the audio-guided saliency splatting strategy. The training objectives are described in Section~\ref{optimize}.

\subsection{Preliminary: 3D Gaussian Splatting}
\label{pre}
Before introducing the proposed Dynamic Gaussian Splatting for Talking Head synthesis, that's take a glance at the preliminary 3D Gaussian Splatting~\cite{kerbl20233d}. 3D Gaussian Splatting explicitly represents the scene as a set of 3D point clouds modeled as Gaussian distribution. A 3D Gaussian for a 3D coordinate $x\in \mathbb{R}^3$ can be defined by a covariance matrix $\Sigma$ and a mean position $\mu$ as,
\begin{equation}\label{eq:1}
G(x)=e^{-\frac{1}{2}(x-\mu)^T\Sigma^{-1}(x-\mu)}.
\end{equation}
To maintain the positive semi-definite of the covariance matrix $\Sigma$ during training optimization, the $\Sigma$ is further decomposed into a scaling matrix $S$ and a rotation matrix $R$, as $\Sigma=RSS^TR^T$. And $S$ and $R$ are optimized with two learnable vectors $s\in \mathbb{R}^3$ for scaling and $r \in \mathbb{R}^4$ for rotation. For modeling of the appearance, each Gaussian also contains the spherical harmonics (SH) and opacity $\alpha$ information. In summary, a 3D Gaussian consists of the following parameters,
\begin{equation}
G=\{\mu, r, s, SH, \alpha\}.
\label{params_GS}
\end{equation}
To project the 3D Gaussians to 2D image space for rendering, the $\Sigma$ can be transferred to the 2D covariance matrix $\Sigma'$ as $\Sigma'=JW\Sigma W^TJ^T$, where $W$ is the viewing transformation and $J$ is the affine approximation Jacobian of the projective transformation. On this basis, the color of each image pixel is computed by blending all overlapping Gaussians, sorted by their opacity, as,
\begin{equation}\label{eq:4}
C=\sum_{i\in N}c_i{\alpha}'_i\prod_{j=1}^{i-1}(1-{\alpha}'_j),
\end{equation}
where $c_i$ denotes the color of each point that is derived from SH coefficients conditioned on the view direction. ${\alpha}'_i$ is computed as the product of the opacity $\alpha$ and the projected covariance $\Sigma'$. 

For simplicity notation, with a set of 3D Gaussian points and a given view direction $P$, the corresponding 2D image $I$ is rendered with the differentiable splatting-based rasterizer $\mathcal{R}$ as,
\begin{equation}\label{eq:3}
I = \mathcal{R}(G, P).
\end{equation}

By calculating the $\mathcal{L}_{L1} = L1(I,\hat{I})$ loss between the rendered image $I$ and the ground truth $\hat{I}$, the traditional 3D Gaussian Splatting backpropagates the gradient, and iteratively updates the Gaussian parameters in Eq.~(\ref{params_GS}), until these Gaussian points adapt to the geometry of the target scene.

\subsection{Dynamic Gaussian Splatting for Talking Head Synthesis}
\label{DGS}
3D Gaussian Splatting is originally designed for static scene modeling~\cite{kerbl20233d}, where the optimization finally yields a fixed set of 3D Gaussian parameters, making it unsuitable for capturing temporal dynamics~\cite{wu20234d}. For the talking head synthesis task addressed in this work, the positions, colors, and other attributes of the 3D Gaussian points must be dynamically updated in response to the input audio signals, particularly in the highly dynamic mouth region. To this end, we incorporate the audio signal condition into the Gaussian Splatting framework, extending it to support temporally coherent and dynamic facial animation when talking.

\begin{figure}[t]
  \centering
   \includegraphics[width=1\linewidth]{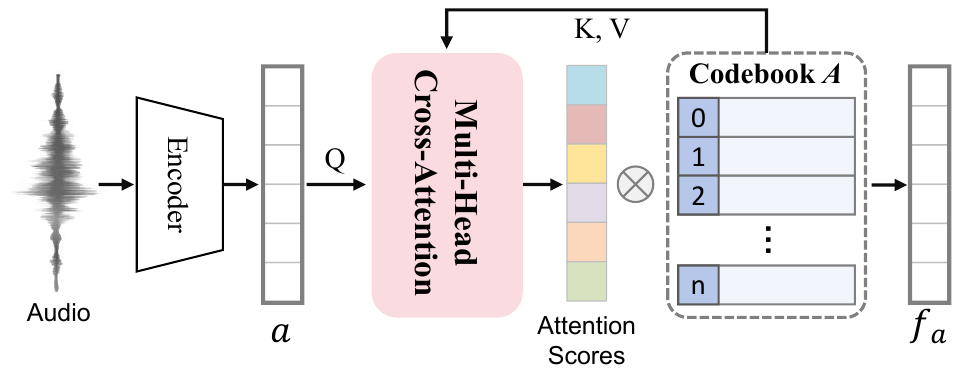}
\vspace{-4mm}
\caption{Illustration of the proposed cross-attention module for audio representation learning. The smoothed audio $a$ acts as the \textit{query}, while the learnable audio codebook $A$ serves as both the \textit{key} and \textit{value}. The resulting attention weights are applied to $A$ to produce the final audio feature $f_a$.}
\label{fig_atten}
\vspace{-2mm}
\end{figure}

To effectively encode the temporal audio information, we utilize a two-stage smoothing operation in the audio encoder. The raw audio signal is first segmented into overlapping windows of 16 time intervals following the practice in VOCA~\cite{cudeiro2019capture}. Then the pre-trained RNN-based DeepSpeech model~\cite{hannun2014deep} is employed to extract per-frame audio features. Adjacent features are further fused into a smoothed audio representation $a\in \mathbb{R}^{D}$ with a learnable temporal filtering~\cite{thies2020neural} for inter-frame consistency, where $D$ is the feature dimension. All these audio processing operations are summed up as the `Encoder' as shown in Fig.~\ref{fig_atten} for simplicity.

Building upon the audio-driven Gaussian Splatting framework, we observe that the dynamics of a talking face are strongly influenced by both the speech signal and the spatial location of facial regions. To capture such audio-conditioned spatial dynamics, we introduce a deformation module $\mathcal{F}$ that learns to dynamically adjust the attributes of each 3D Gaussian point based on both the audio signal and spatial position. Specifically, the module takes as input the audio feature $a$ and the 3D position $(x, y, z)$ of each Gaussian point $G$, and predicts corresponding attribute offsets $\bigtriangleup G$.
This constructs a dynamic 4D representation $G'$ for the talking head, where each Gaussian is temporally modulated based on both audio content and spatial context. The deformation process can be formulated as,
\begin{equation}\label{eq_delta_g}
G' = G + \bigtriangleup  G = G + \mathcal{F}(G, a, x, y, z),
\end{equation}
where $\bigtriangleup  G = \mathcal{F}(G, a, x, y, z)$ represents the learned Gaussian attribute offsets (\emph{e.g.}, position, color, opacity). In this way, our framework effectively bridges the gap between conventional static 3D Gaussian Splatting and the audio-driven dynamic Gaussian Splatting, for realistic and synchronized talking head synthesis. In the following subsection, we will detail the designed deformation module $\mathcal{F}$.

\subsection{Audio Factorization Plane}
\label{Sptial}

With the overall learning objective defined in Eq.~(\ref{eq_delta_g}), the primary challenge now lies in designing the deformation module $\mathcal{F}$ to effectively integrate the audio input $a$ with the spatial coordinates $(x, y, z)$ of each Gaussian point. 
A naive solution is to directly store a dense 4D volume, where the three spatial dimensions are extended with an additional temporal audio dimension. However, this approach is memory-consuming due to the curse of dimensionality. Moreover, since audio features are high-dimensional vectors rather than scalar quantities, it is non-trivial to directly incorporate them as a coordinate axis. Another straightforward solution, as illustrated in Fig.~\ref{fig1} (b), is to extract audio and spatial features independently and concatenate them for subsequent learning. Although such paradigm explicitly models the 3D spatial field, it relies entirely on an MLP to implicitly learn the cross-modal interaction between audio and geometry, which limits its interpretability and the ability to infer fine-grained and structured audio-conditioned facial dynamics.
To address this limitation, we propose an explicit representation of the audio-driven 4D volume through structured audio factorization, which enables more interpretable and spatially aware modeling of speech-driven facial deformation.

To this end, we propose the Audio Factorization Plane representation (For simplicity, we also refer to it as `Audio-Plane' in the following discussion), which explicitly encodes the audio-driven 4D deformation field through a set of structured and interpretable planes for audio-driven talking head modeling. Instead of treating audio as a global condition and concatenating it with the spatial coordinates directly, we factorize the 4D volume into two complementary components: (1) audio-independent spatial planes as $P_{XY}, P_{XZ}, P_{YZ}$, and (2) audio-conditioned modulation planes as $P_{ZA}, P_{YA}, P_{XA}$, just as shown in Fig.~\ref{fig2}. The former three planes capture the static geometric structure of the face, while the latter three audio-conditioned planes introduce fine-grained, speech-dependent variations across spatial dimensions.

Specifically, the spatially correlated feature planes $P_{XY}, P_{XZ}, P_{YZ}$, each with the dimension of $\mathbb{R}^{H\times H\times D}$, follow the traditional tri-plane spatial representation~\cite{chan2022efficient}, where $H$ denotes the grid resolution and $D$ the feature dimension. This tri-plane effectively encodes static spatial information for 3D geometry. Building upon this foundation, we introduce our core contribution: audio-spatially correlated planes $P_{ZA}, P_{YA}, P_{XA}$, which enable the model to capture dynamic facial variations conditioned jointly on spatial location and audio content.
A key challenge in constructing these audio-dependent planes lies in the fact that audio features are high-dimensional vectors rather than scalar quantities, making it non-trivial to embed them directly as coordinate dimensions. To address this, we propose to factorize each of the audio-dependent planes $P_{ZA}, P_{YA}, P_{XA}$ into a pair of interacting feature grids: a shared learnable audio prototype grid $A \in \mathbb{R}^{H \times D}$ and one of the three spatial feature grids $L_{X}, L_{Y}, L_{Z} \in \mathbb{R}^{H\times D}$. It needs to be emphasized that the audio prototype grid $A$ is shared across all three audio-conditioned planes. The audio prototype grid serves as a latent codebook that captures prototypical phoneme-level patterns, enabling the model to modulate spatial deformation in a structured and semantically meaningful manner.
The three spatial grid embeddings $L_{X}, L_{Y}, L_{Z}$ each correspond to one spatial dimension. 
These paired feature grids are then employed to span the three audio-spatially correlated feature planes as illustrated in Fig.~\ref{fig2}, as,
\begin{equation}
\{L_{X}, A\}\Rightarrow P_{XA},\:\{L_{Y}, A\}\Rightarrow P_{YA},\:\{L_{Z}, A\}\Rightarrow P_{ZA}.
\end{equation}
Here, the symbol `$\Rightarrow$' is used to indicate the plan span operation.

In the following, we provide a detailed explanation of how the audio prototype grid $A$ interacts with the spatial grids to span the Audio Factorization Plane, and how this representation is used to compute the final audio-conditioned feature for a specific 3D Gaussian point.
For a given 3D Gaussian point $(x,y,z)$ under an input audio condition $a\in \mathbb{R}^{D}$, we design a cross-attention module to obtain its audio representation $f_{a} \in \mathbb{R}^{D}$ from the audio feature grid $A\in \mathbb{R}^{H \times D}$, as shown in Fig.~\ref{fig_atten}. In this attention module, the input smoothed audio $a$ acts as the \textit{query} information, while the audio grid $A$ serves as both the \textit{key} and \textit{value}. The specific cross-attention computation is formulated as,
\begin{equation} 
f_{a} = \text{Attention}(Q,K,V)=\text{Softmax}\left ( \frac{QK^{T}}{\sqrt{d_{k}}} \right )V,
\label{attention}
\end{equation}
where Q, K, V are queries, keys, and values, with $K=V=A$ and $Q=a$, and $\sqrt{d_{k}}$ is a scaling factor that stabilizes gradients. 
In parallel, the spatial embedding of this Gaussian point on the spatial grid $L_{Z}$ is obtained via bilinear interpolation, resulting in a spatial feature vector $f_{z} \in \mathbb{R}^{D}$. On this basis, the combined audio-spatial feature on the corresponding factorized plane $P_{ZA}$ is then formulated as,
\begin{equation}
f_{za} = f_z \cdot f_a,
\end{equation}
where `$\cdot$' denotes element-wise multiplication.
This process is repeated similarly for the other two audio-spatial planes $P_{XA}$ and $P_{YA}$, using the spatial grids $L_{X}$ and $L_{Y}$ respectively.
In this way, the audio feature grid $A$ pairing with each spatial feature grid $L_{X}, L_{Y}, L_{Z}$ can span the audio-spatial correlation feature planes. It's worth noting that although we refer to $P_{ZA}, P_{YA}, P_{XA}$ as `planes', they are in fact spanned implicitly through the interaction of audio and spatial feature grids, rather than being explicitly stored as 2D tensors. We therefore visualize them with dashed lines in Fig.~\ref{fig2} for illustration. 
This factorized design allows the model to capture fine-grained temporal-spatial dependencies using a compact representation, significantly enhancing its ability to model dynamic audio-driven lip movements.

In practice, we adopt multi-resolution feature planes to capture both coarse and fine-grained audio-spatial correlations. We denote the complete Audio Factorization Plane representation as $\mathcal{A}$, which consists of the following components,
\begin{equation}
\mathcal{A} = \{P^{s}_{XY}, P^{s}_{XZ}, P^{s}_{YZ}, L^{s}_{X}, L^{s}_{Y}, L^{s}_{Z}, A^{s}\}, 
\end{equation}
where $s$ is the upsampling scale, with $P^{s}_{XY} \in \mathbb{R}^{sH\times sH\times D}$, $L^{s}_{X} \in \mathbb{R}^{sH\times D}$, $A^{s} \in \mathbb{R}^{sH\times D}$ and so on. On this basis, the process for computing feature $f$ of a Gaussian point from the proposed Audio-Plane can be formulated as, 
\begin{equation}\label{eq_delta}
f = \bigcup_{(s)} \bigcup_{(i,j,k)}(\text{Interp}(P^{s}_{ij})\cdot (f^{s}_k)\cdot f^{s}_a),
\end{equation}
where $(i,j,k)\in \left \{ (X, Y, Z), (X, Z, Y), (Y, Z, X) \right \}$, $s\in \left \{ 1,2,3,4 \right \}$. `Interp' indicates the bilinear interpolation operation used to sample spatial features. `$\cdot$' refers to element-wise multiplication for feature modulation. The symbol `$\bigcup$' represents the concatenation of features along the channel dimension.

Based on the learned audio-aware spatial feature $f$, we then employ an MLP network $\mathcal{D}_1$ as shown in Fig.~\ref{fig2} to predict the Gaussian deformation as,
\begin{equation}
\bigtriangleup  G = \mathcal{D}_1(f).
\end{equation} 
This enables the 3D Gaussians to be dynamically modulated by the input audio signal, thereby extending the static representation to support audio-driven dynamic talking head modeling.

\subsection{Audio-Guided Saliency Splatting}
\label{weight}

With the Audio-Plane introduced above, we couple audio with spatial information to guide the dynamic learning of Gaussian points. 
However, the current deformation strategy treats all facial regions uniformly, without accounting for the fact that different face regions exhibit varying degrees of motion dynamics. In practice, areas such as the mouth have more frequent and complex deformations during speech, while other regions like the cheeks or forehead, remain relatively static.
This motivates us to design a region-aware deformation field that assigns different dynamic importance to different facial areas. In particular, the model is encouraged to focus on learning rich and expressive deformations in highly dynamic regions (\emph{e.g.} the mouth), while suppressing redundant motion in more static regions, which contributes to improved generation realism and temporal stability.
Based on these characteristics, we refer to this adaptive splatting strategy as `saliency splatting'.

To this end, we propose to assign a dynamic attribute $d$ to each Gaussian point, which controls the degree of dynamic variation according to its spatial and audio context. As shown in Fig.~\ref{fig2}, this attribute $d\in \mathbb{R}^{1}$ is predicted from the audio-aware spatial feature $f$ with a lightweight MLP network $\mathcal{D}_2$ as,
\begin{equation}
d = \mathcal{D}_2(f).
\end{equation}
Let $D=\{d\}$ denote the set of dynamic weights for all Gaussians. The final deformed Gaussian representation $G'$ is then computed as,
\begin{equation}
G' = G+D\cdot \bigtriangleup G.
\end{equation}

Then the challenge is how to supervise the learning of $d$ to achieve the desired region-aware dynamic modulation.
Due to the unordered nature of the spatial Gaussian points, it is non-trivial to directly determine which specific facial region each Gaussian point belongs to. To address this, we propose the audio-guided saliency splatting mechanism as illustrated in Fig.~\ref{fig2}.
Specifically, the learned dynamic weights $D=\{d\}$ are projected back into the 2D image space under view direction $P$ through the differentiable splatting-based rasterizer $\mathcal{R}$, producing a grayscale dynamic map $g$ with the same resolution as the face image as,
\begin{equation}
g = \mathcal{R}(D, P).
\end{equation}
Each pixel in $g$ reflects the accumulated dynamic influence of Gaussian points projected to that location, thereby providing a spatially continuous representation of motion intensity.

To guide the learning of these dynamic weights, we design reference masks $m$ as shown in Fig.~\ref{fig2}, which serve as supervisory signals for the distribution of dynamic regions. These masks are constructed by applying a dilation operation based on the mouth region mask, capturing an extended area of expected motion. By constraining the similarity between the predicted dynamic map $g$ and the reference mask $m$, we explicitly guide the model to assign higher dynamic weights to Gaussian points near the mouth region, where motion is typically more intense during speech.
Furthermore, since the reference mask $m$ is a binary map representing an absolute $0\text{-}1$ distribution, whereas the actual dynamism weights are continuous values, we introduce a margin-based loss $\mathcal{L}_\text{mask}$ to allow for flexible alignment between the two as,
\begin{equation}
\mathcal{L}_\text{mask}=\text{max}(0, \left \| m-g \right \| - \text{margin}).
\label{eq_margin}
\end{equation}
This relaxed strategy allows the dynamism weight $d$ to adjust adaptively within a permissible range, rather than being forced to match a hard binary distribution.

The proposed audio-guided saliency splatting mechanism encourages the model to focus more effectively on the highly dynamic mouth region, while suppressing unnecessary deformation in relatively static facial areas. This modeling strategy improves representation efficiency and enhances the accuracy of audio-lip synchronization. Moreover, this region-aware supervision provides structural guidance that facilitates more stable and efficient training.

\subsection{Training Objectives}
\label{optimize}

We adopt a two-stage training strategy, following the practice in 3DGS, consisting of a coarse and a fine training phase. In the coarse training stage, facial dynamics are not yet modeled. The objective at this stage is to rapidly fit the foundational 3D geometry and optimize image-level reconstruction quality, through minimizing an $L1$ loss and a perceptual similarity loss (LPIPS)~\cite{zhang2018unreasonable} between the rendered images and the ground truth as,
\begin{equation}
\mathcal{L}_\text{coarse} = \mathcal{L}_{L1} + \lambda_\text{LPIPS} \mathcal{L}_\text{LPIPS}.
\label{all_loss1}
\end{equation}
This coarse stage provides a good initialization for the 3D Gaussian points, enabling them to roughly fit the basic facial structure and view-dependent appearance. Building on this foundation, in the fine training phase, we introduce the Audio-Plane and the audio-guided saliency splatting mechanism to model the audio-driven facial dynamics with region-aware modulation. To guide the learning of the deformation field, we incorporate the additional region-aware loss $\mathcal{L}_\text{mask}$, which encourages the model to focus on highly dynamic regions such as the mouth.
The overall objective in the fine stage is defined as,
\begin{equation}
\mathcal{L}_\text{fine} = \mathcal{L}_{L1} + \lambda_\text{LPIPS} \mathcal{L}_\text{LPIPS} + \lambda_\text{mask} \mathcal{L}_\text{mask}.
\label{all_loss2}
\end{equation}

\section{Experiments}

\subsection{Experimental Settings}
\textbf{Dataset. }
We use the video clips publicly released in some previous methods~\cite{guo2021ad,shen2022dfrf} as the datasets for experiments. These high-resolution videos are captured in real life scene with an average length of 4 minutes, allowing us to evaluate the performance of talking head generation methods in practical application scenarios. We set the video Frame Per Second (FPS) to 25 and the resolution to $512\times 512$ for high-resolution synthesis. The head pose is estimated through the off-line face tracking tool developed in~\cite{thies2016face2face}. 
For each target subject, we use an approximately 80-second video clip for training, which corresponds to roughly 2000 frames at 25fps, while reserving other non-overlapping clips for inference.

\vspace{1mm}
\textbf{Metric.}
We assess the performance of our proposed method using a combination of visual results and quantitative indicators. The primary metrics for evaluating image quality include PSNR ($\uparrow$), SSIM ($\uparrow$)~\cite{wang2004image} and LPIPS ($\downarrow$)~\cite{zhang2018unreasonable}. PSNR and SSIM are traditional measures that assess the fidelity and structural similarity of the generated images compared to the ground truth. LPIPS is a learning-based perceptual similarity measure that aligns more closely with human perception and is recommended as a more objective indicator. To evaluate the audio-visual synchronization quality, which is crucial for the audio-driven talking head generation task, we use the SyncNet Confidence score ($\uparrow$)~\cite{chung2016out}. In our notation, a higher score `$\uparrow$' indicates better performance for PSNR, SSIM, and SyncNet Confidence, while a lower score `$\downarrow$' is preferred for LPIPS.

\vspace{1mm}
\textbf{Implementation Details.}
In the coarse training phase, we randomly initialize 1,000 Gaussian points and train for 3,000 iterations to optimize the Gaussian points. In the fine training stage, we jointly optimize both the Gaussian points, the Audio-Plane, and the deformation field for 100,000 iterations. The dimension of each feature in the space feature planes $P_{XY}, P_{XZ}, P_{YZ}$ is $D=16$, and the grid resolution of these feature planes is set as $H=64$. The loss weights used in Eq.~(\ref{all_loss1}) and (\ref{all_loss2}) are configured as $\lambda_\text{LPIPS}=0.05$ and $\lambda_\text{mask}=0.1$, and the margin of the $\mathcal{L}_\text{mask}$ in Eq.~(\ref{eq_margin}) is set as 0.2, which are selected according to the experiments. During inference, the synthesized head part is combined with the torso part and background to produce the final result. All experiments are conducted on one NVIDIA 3090 Ti GPU. 

\subsection{Ablation Study}
To better evaluate the contribution of key components and design choices in our framework, we conduct a series of ablation studies.
In particular, we examine the impact of the Audio-Plane representation, the audio-guided saliency splatting mechanism, and explore alternative designs of the Audio-Plane. We also investigate the effect of different margin values in the region-aware loss $L_{mask}$.
By selectively disabling or modifying these components, we analyze how each design affects the quality of the generated talking head videos.

\begin{table}[tb]
\centering
\caption{Ablation study on the contribution of the proposed Audio-Plane (`AP') and the Saliency Splatting (`SS').}
\renewcommand\tabcolsep{9pt}
\begin{tabular}{c c c c c} 
\toprule
Method&PSNR$\uparrow$ &SSIM$\uparrow$ & LPIPS$\downarrow$  & SyncNet$\uparrow$ \\
\midrule
GT&-&-&-&9.175\\
\midrule
w/o AP \& SS &30.73&0.920&0.041&4.565\\
+ SS&30.81&0.923&0.039&5.427\\
+ AP&31.54&0.925&0.037&6.342\\
\midrule
+ AP \& SS (Ours)&\textbf{31.69}&\textbf{0.929}&\textbf{0.035}&\textbf{6.700}\\
\bottomrule
\end{tabular}
\vspace{-2mm}
\label{table1}
\end{table}

\vspace{1mm}
\textbf{Effect of the Audio-Plane and Saliency Splatting.}
In this section, we investigate the effect of the proposed Audio Factorization Plane (denoted as `AP') and the audio-guided Saliency Splatting mechanism (denoted as `SS'), as shown in Table~\ref{table1}. In the `w/o AP \& SS' baseline, the audio feature is directly concatenated with spatial embedding to drive a Gaussian Splatting model, following the structure illustrated in Fig.~\ref{fig1} (b). As shown in the second row of Table~\ref{table1} (`+ SS'), introducing the saliency splatting method boosts the LPIPS score to 0.039 and the SyncNet score to 5.427, indicating better perceptual quality and audio-lip synchronization. This improvement is attributed to the region-aware deformation field, which enables finer modeling of motion in dynamic facial regions. Further, the results in the third row (`+ AP') show that incorporating the Audio-Plane enhances LPIPS to 0.037 and SyncNet to 6.342. This suggests that the compact and structured audio-spatial representation provided by the Audio-Plane facilitates more coherent and synchronized lip movements.
Finally, combining both the saliency splatting and the Audio-Plane (`+ AP \& SS') yields superior results, highlighting the complementary benefits of compact audio-spatial representation and region-aware deformation field in our framework.

\begin{figure*}[t]
  \centering
   \includegraphics[width=1\linewidth]{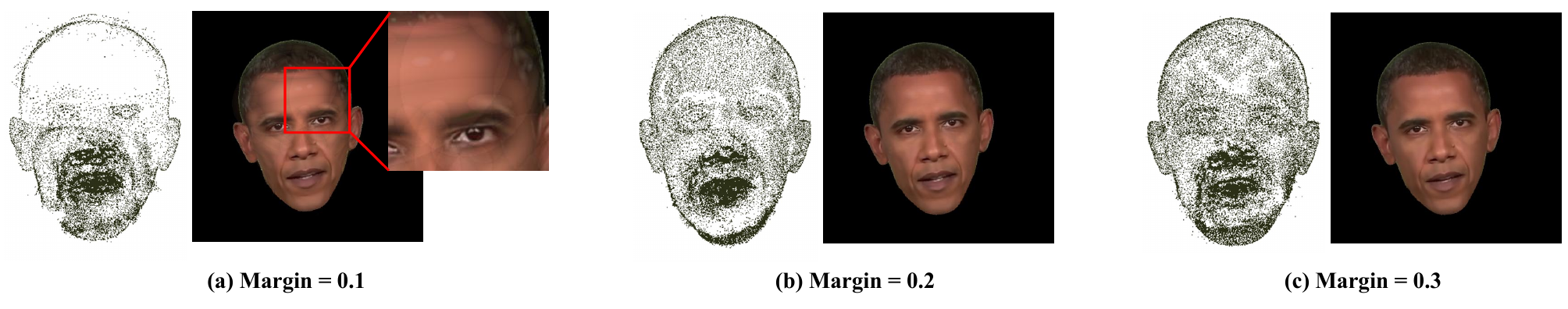}
\vspace{-4mm}
\caption{Generated frames and the corresponding Gaussian points under different margins. For intuitive comparisons, we only show the head area. Results show that a moderate margin as 0.2 yields more dense Gaussians in the highly deformable mouth region for accurate lip motion modeling. A low margin as 0.1 causes too sparse Gaussian points and artifacts, while a too high margin as 0.3 weakens focus on high dynamic regions by over-smoothing the Gaussian distribution.}
\label{fig3}
\end{figure*}

\begin{table}[t]
\centering
\caption{Comparison of different Audio-Plane designs across two backbone variants using various fusion methods.}
\renewcommand\tabcolsep{9pt}
\begin{tabular}{c c c c c} 
\toprule
Backbone&Fusion&PSNR$\uparrow$ & LPIPS$\downarrow$  & SyncNet$\uparrow$  \\
\midrule
GT&-&-&-&8.266\\
\midrule
\multirow{3}{*}{MLP}&multiply&31.43&0.093&4.447\\
&add&32.10&0.043&4.915\\
&concat&32.22&0.043&5.149\\
\midrule
\multirow{3}{*}{\makecell[c]{Cross- \\ Attention}}&multiply&31.65&0.052&5.170\\
&add&32.37&0.042&5.230\\
&concat&\textbf{32.88}&\textbf{0.038}&\textbf{5.518}\\
\bottomrule
\end{tabular}
\vspace{-2mm}
\label{tablea}
\end{table}

\vspace{1mm}
\textbf{Different Designs in Audio-Plane.}
In the proposed Audio-Plane, we design a cross-attention module between the input audio and the learnable audio prototype grid, as illustrated in Fig.~\ref{fig_atten}.
To assess the effectiveness of this design, we conduct a comparison by replacing the attention module with a simple MLP that directly predicts the attention scores, removing the explicit query-key-value interaction.
Additionally, since the Audio-Plane comprises three sets of feature planes (\textit{e.g.} each set includes a spatial plane such as $P_{XY}$ and its corresponding audio-conditioned plane such as $P_{ZA}$), the feature fusion strategy used to combine these planes plays an important role in the final performance. Here we explore three fusion strategies including element-wise multiplication (`\textit{Multiply}'), element-wise addition (`\textit{Add}'), and channel-wise concatenation (`\textit{Concatenate}'). Table~\ref{tablea} presents the results across two backbone variants using these different fusion methods.
Among all configurations, the cross-attention backbone combined with concatenation fusion consistently achieves the best performance.
This superior result can be attributed to two key factors. 
First, the cross-attention mechanism enables the model to dynamically select relevant audio tokens based on spatial context, allowing for more precise alignment between phonetic content and localized facial motion.
Second, the concatenation-based fusion strategy preserves the full information from both spatial and audio streams, providing richer and more flexible feature representations than addition or multiplication fusion.
Together, this configuration facilitates more precise modeling of audio-conditioned facial dynamics, leading to improvements in both perceptual quality and audio-lip synchronization. These results guide us to select the cross-attention module with concatenation fusion mode in the following experiments.

\begin{table}[t]
\centering
\caption{Impact of different selections of margin in $\mathcal{L}_\text{mask}$. Setting a moderate margin as 0.2 yields the comprehensive optimal result, balancing both the perceptual quality and the audio-lip synchronization.}
\renewcommand\tabcolsep{11pt}
\begin{tabular}{c c c c c} 
\toprule
Margin&PSNR$\uparrow$ &SSIM$\uparrow$ & LPIPS$\downarrow$  & SyncNet$\uparrow$  \\
\midrule
GT&-&-&-&6.915\\
\midrule
0.05&31.76&0.943&0.062&4.443\\
0.1&32.02&0.950&0.047&4.760\\
0.2&31.95&\textbf{0.954}&\textbf{0.042}&\textbf{5.176}\\
0.3&\textbf{32.20}&0.952&\textbf{0.042}&4.615\\
\bottomrule
\end{tabular}
\vspace{-2mm}
\label{tableb}
\end{table}

\vspace{1mm}
\textbf{Investigation on Margin Selections for the Saliency Splatting.}
In the design of the audio-guided saliency splatting, we use mouth mask images as supervision to guide region-aware deformation. However, the actual dynamic weights are continuous values and do not strictly follow a binary $0\text{-}1$ distribution. We therefore present the margin-based loss $\mathcal{L}_\text{mask}$ defined in Eq.~(\ref{eq_margin}), which introduces a tolerance between the predicted dynamic map and the binary mask. Here we investigate the impact of different margin values on model performance. Quantitative results in Table~\ref{tableb} illustrate that setting the margin as 0.2 yields the comprehensive optimal result, balancing both perceptual quality and audio-lip synchronization.
To better understand the effect of the margin, we also visualize the generated face images and their corresponding Gaussian point clouds under different margin settings in Fig.~\ref{fig3}. It can be seen that with the audio-guided saliency splatting strategy, there are more dense Gaussian points in the mouth region, indicating the model’s focus on highly dynamic areas.
However, when the margin is too low (e.g., Margin = 0.1 in Fig.~\ref{fig3} (a)), the supervision becomes overly strict, resulting in sparser point distributions and visible artifacts in the synthesized face, as highlighted in the zoomed-in view where the edge of the Gaussian can be seen.
Conversely, setting the margin too high (e.g., Margin = 0.3 in Fig.~\ref{fig3} (c)) causes the dynamic weights to spread more uniformly across the face, reducing focus on the mouth region and weakening lip motion modeling.
Based on both quantitative results and qualitative observations, we adopt an appropriate margin value of 0.2 in the following experiments.

\begin{figure*}[t]
  \centering
   \includegraphics[width=1\linewidth]{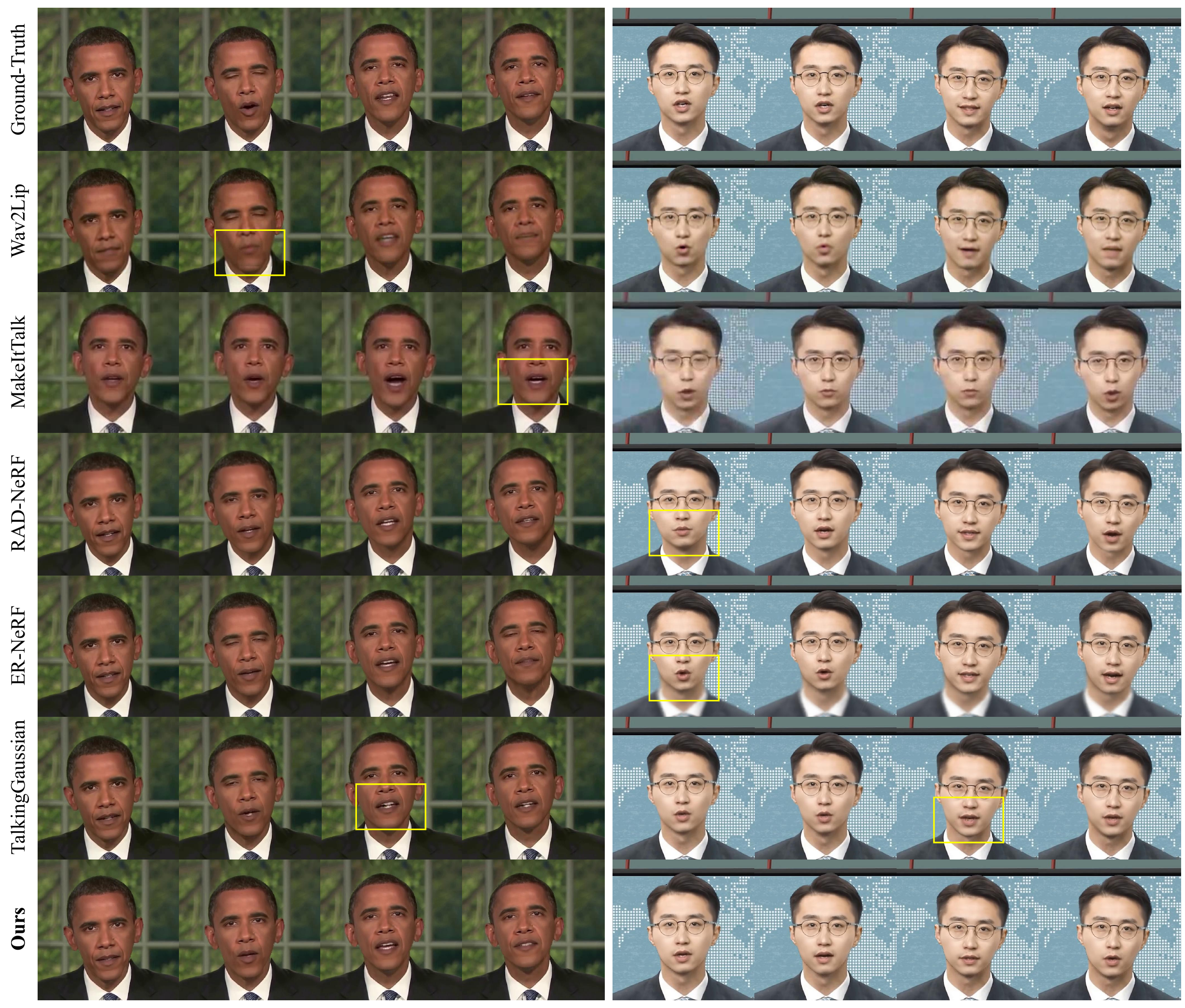}
\vspace{-4mm}
\caption{Visual comparison with some representative talking head synthesis methods, including 2D-based ones Wav2lip~\cite{prajwal2020lip} and MakeItTalk~\cite{zhou2020makelttalk}, NeRF-based ones RAD-NeRF~\cite{tang2022real} and ER-NeRF~\cite{li2023efficient}, and Gaussian Splatting-based approach TalkingGaussian~\cite{li2024talkinggaussian}.
We used yellow boxes to highlight some noticeable artifacts or inaccurate mouth shapes.
Our method, empowered by the proposed Audio-Plane and audio-guided saliency splatting strategy, produces more faithful facial animations with fewer artifacts and improved audio-lip synchronization.}
\label{fig4}
\vspace{-2mm}
\end{figure*}

\begin{table*}[t]
\centering
\caption{
Comparison with some representative talking head synthesis methods on two test sets. We show the best and second-best results with \textbf{blod} and \underline{underline}, respectively. Our method achieves real-time rendering speed with comprehensive optimal results on both visual quality and audio-visual synchronization metrics.}
\renewcommand\tabcolsep{10pt}
\begin{tabular}{c c c c c c c c c c} 
\toprule
\multirow{2}{*}{Method} & \multicolumn{4}{c}{Test Set A}  &\multicolumn{4}{c}{Test Set B}& \multirow{2}{*}{FPS}\\
\cmidrule(lr{0.6em}){2-5}
\cmidrule(lr{0.6em}){6-9}
&PSNR$\uparrow$ &SSIM$\uparrow$ & LPIPS$\downarrow$& SyncNet$\uparrow$  & PSNR$\uparrow$ &SSIM$\uparrow$ & LPIPS$\downarrow$& SyncNet$\uparrow$\\
\midrule
GT &-&-&-&9.175& -&- &-&8.266&-\\
\midrule
ATVG~\cite{chen2019hierarchical}& 29.50 &0.907 &0.057 &4.152& 28.98&0.899&0.104&2.852& \\
MakeItTalk~\cite{zhou2020makelttalk}&29.46&0.908&0.092&5.647&31.75&0.945&0.042&4.906&10\\
Wav2Lip~\cite{prajwal2020lip}&31.58&0.913&0.082&\textbf{8.554}&32.58&0.951&0.092&\textbf{8.044}&15\\
\midrule
AD-NeRF~\cite{guo2021ad}&\underline{31.65}&0.922&0.082&4.311&32.33&0.953&0.048&4.321&0.1\\
RAD-NeRF~\cite{tang2022real}&30.20&\underline{0.927}&0.041&6.537&\underline{32.86}&	0.946&\underline{0.039}&5.323&32\\
ER-NeRF~\cite{li2023efficient}&29.66&0.921&\underline{0.037}&6.549&32.83&0.944&0.041&5.124&34\\
SyncTalk~\cite{peng2024synctalk}&31.28&\underline{0.927}&\textbf{0.035}&6.531&32.77&\underline{0.954}&\underline{0.039}&5.316&28\\
\midrule
GaussianTalker~\cite{cho2024gaussiantalker}&29.05&0.913&0.042&6.438&30.16&0.922&0.043&5.059&20\\
TalkingGaussian~\cite{li2024talkinggaussian}&28.92&0.910&0.043&5.026&30.83&0.923&0.042& 5.405&\textbf{45}\\
\midrule
\textbf{Ours}&\textbf{31.69}&\textbf{0.929}&\textbf{0.035}&\underline{6.700}&\textbf{32.88}&\textbf{0.956}&\textbf{0.038}&\underline{5.518}&\underline{43}\\
\bottomrule
\end{tabular}
\label{table4}
\end{table*}

\subsection{Method Comparison}
In this section, we compare our proposed method with several representative talking head synthesis approaches, spanning multiple categories as: 2D-based methods ATVG~\cite{chen2019hierarchical}, MakeItTalk~\cite{zhou2020makelttalk} and Wav2Lip~\cite{prajwal2020lip}; NeRF-based approaches AD-NeRF~\cite{guo2021ad}, RAD-NeRF~\cite{tang2022real}, ER-NeRF~\cite{li2023efficient} and SyncTalk~\cite{peng2024synctalk}; and more recent Gaussian Splatting-based ones GaussianTalker~\cite{cho2024gaussiantalker} and TalkingGaussian~\cite{li2024talkinggaussian}.
We perform method comparison under two experimental protocols: self-driven and cross-driven settings. In the self-driven setting, the driving audio is from the same identity as the test subject, allowing for intra-identity evaluation of audio-lip synchronization accuracy and appearance fidelity. In contrast, the cross-driven setting uses audio from a different identity to assess the model’s ability to generalize across speakers.

\vspace{1mm}
\textbf{Self-Driven Results.}
In the self-driven setting, we conduct evaluations on two test sets, using standard metrics for visual quality (PSNR, SSIM, and LPIPS), audio-visual synchronization (SyncNet score), and inference speed (frames per second, FPS).
Comprehensive quantitative comparisons are shown in Table~\ref{table4}, where we highlight the best and second-best results with
\textbf{blod} and \underline{underline}, respectively.
Our method consistently achieves the best or second-best scores across all metrics, demonstrating its comprehensive superiority in generating high-quality, synchronized, and real-time talking head videos.
It can be seen that our method achieves the best LPIPS score of 0.035 on Test Set A, and 0.038 on Test Set B, indicating superior perceptual quality.
In terms of SyncNet, our method ranks second, just behind Wav2Lip. 
However, while Wav2Lip achieves the best SyncNet score, it exhibits significantly lower LPIPS values, suggesting poor reconstruction quality. This can also be clearly observed in the supplementary video. Moreover, our approach achieves a rendering speed of 43 FPS, exceeding the 25 FPS requirement for real-time playback. This highlights its practical potential for real-time talking head applications.

For better visualization, Fig.~\ref{fig4} shows the visual comparison with some representative baselines.
We used yellow boxes to highlight some inaccurate mouth shapes or artifacts.
Our method, powered by the Audio-Plane and region-aware modeling, produces facial animations that closely align with ground-truth video frames, exhibiting natural motion and precise lip synchronization.
In comparison, 2D-based methods such as MakeItTalk~\cite{zhou2020makelttalk} and Wav2Lip~\cite{prajwal2020lip} suffer from shape distortions and lower visual fidelity with some mismatch in the mouth shape.
NeRF-based methods like RAD-NeRF~\cite{tang2022real} and ER-NeRF~\cite{li2023efficient} produce relatively high-quality frames, but still produce inaccurate lip movements in several cases. TalkingGaussian~\cite{li2024talkinggaussian}, which separately models the mouth and head regions, introduces visual inconsistencies and artifacts around the teeth area due to component misalignment.
More results and comparisons can be found in our supplementary video, which clearly shows the advantages of our method in both fidelity and temporal coherence.

\begin{table}[tb]
\begin{center}
\centering
\caption{Method comparison under the cross-driven setting. These two test audio sets are from the demo of SynObama~\cite{suwajanakorn2017synthesizing} and Neural Voice Puppetry (NVP)~\cite{thies2020neural}, respectively.}
\label{table_cross}
\renewcommand\tabcolsep{8pt}
\begin{tabular}{c c c c c} 
\toprule
\multirow{2}{*}{Method} & \multicolumn{2}{c}{Test Audio A} &\multicolumn{2}{c}{Test Audio B} \\
\cmidrule(rl{0.6em}){2-3}
\cmidrule(rl{0.6em}){4-5}
& LPIPS$\downarrow$ & SyncNet$\uparrow$ & LPIPS$\downarrow$ & SyncNet$\uparrow$\\
\midrule
SynObama~\cite{suwajanakorn2017synthesizing} &-& 4.301&-&-\\
NVP~\cite{thies2020neural} &-&-&-&4.677 \\
ER-NeRF~\cite{li2023efficient}&0.048&4.651&0.045&4.779\\
TG~\cite{li2024talkinggaussian}&0.043&4.372&0.043&4.423\\
\midrule
Ours&\textbf{0.041}&\textbf{4.981}&\textbf{0.042}&\textbf{4.884}\\
\bottomrule
\end{tabular}
\vspace{-3mm}
\end{center}
\end{table}

\vspace{1mm}
\textbf{Cross-Driven Results.}
In the cross-driven setting, the input audio originates from different identities not seen during the model’s training. To evaluate the generalization ability of our method under such mismatched conditions, we use two publicly available audio samples, with the Test Audio A from the SynObama demo~\cite{suwajanakorn2017synthesizing} and the Test Audio B from the Neural Voice Puppetry (NVP) demo~\cite{thies2020neural}.
Although these two methods are not publicly released, we include them in our evaluation by extracting their generated videos directly from their official demo.
As reported in Table~\ref{table_cross}, our proposed model demonstrates strong generalization capability, maintaining high visual fidelity and accurate audio-lip synchronization even when driven by audio from different speakers. This robustness makes our method well-suited for practical scenarios where a single identity model must respond to diverse driving voices, such as in dubbing, voice replacement, or avatar-based communication systems.

\begin{table}[t]
\renewcommand\tabcolsep{7pt}
\begin{center}
\caption{SyncNet scores under the cross-language driven setting. Models trained on three different subjects are evaluated using audio clips from both same-identity and cross-language sources to assess generalization across languages.}
 \label{crosslan}
 \begin{tabular}{c c  c c c c} 
     \toprule
     Driven & Same & English  & Chinese  & French &German \\
      Audio& Id. & (M) &  (M) & (F) &(F) \\
     \midrule
     English (M) &5.518&4.900&4.499&4.344&4.980\\ 
     Chinese (M) &5.895&4.605&5.216&3.741&4.812\\ 
     French (M) &5.143&3.460&4.033&5.017&4.069\\ 
     \bottomrule
 \end{tabular}
 \vspace{-2mm}
 \end{center}
\end{table}

\subsection{Cross-Language Driven Results}
To demonstrate the robustness of our method under the cross-language audio driving setting, we conduct experiments using speech inputs from various languages.
While this experiment does not include comparisons with other methods (which have been done as Table~\ref{table4} and Table~\ref{table_cross} in the previous subsection), its primary objective is to validate the generalization ability of our model when driven by unseen and linguistically diverse audio inputs.
Thanks to the design of our audio prototype grid, our approach naturally supports adaptation to new driving audio.
As shown in Table~\ref{crosslan}, we train models of three different subjects (vertical table headers), and evaluate them using audio clips from both the same identity and four additional languages spoken by different identities (listed in the horizontal table headers).
In addition to achieving the best performance when driven by same-identity audio as expected, our method also yields satisfactory SyncNet scores across various source-target language pairs, demonstrating robust audio-visual synchronization even in cases with significant linguistic and phonetic differences. The corresponding visual results can be more intuitively observed in the video demonstration we provided.
These results highlight the robustness and generalization capability of our method, making it well-suited for multilingual and speaker-agnostic talking head applications.

\begin{figure*}[ht]
  \centering
   \includegraphics[width=1\linewidth]{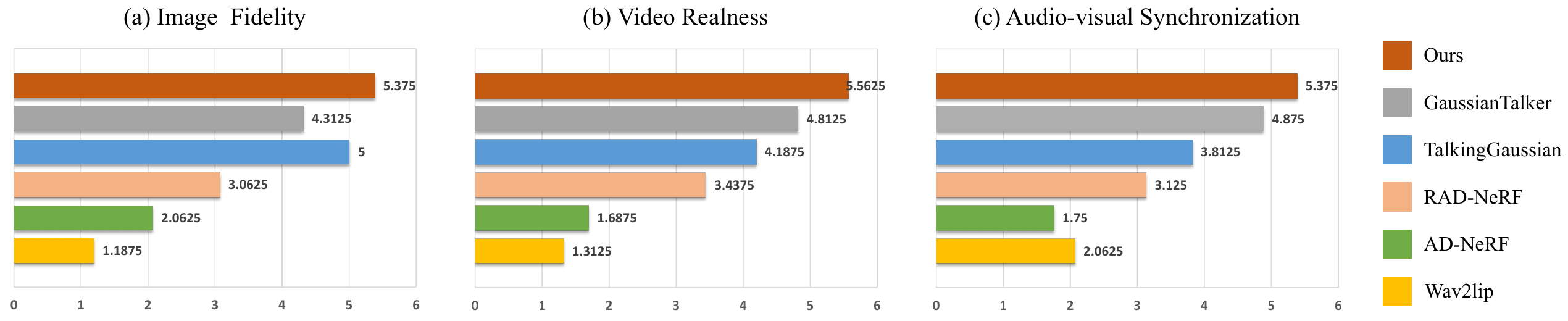}
\vspace{-3mm}
\caption{User study results across some representative talking head synthesis methods on three aspects, \emph{i.e.} \emph{Image Fidelity}, \emph{Video Realness}, and \emph{Audio-visual Synchronization}. It can be seen that, from the subjective human perspective, our method consistently outperforms other baselines across all evaluation dimensions.}
\label{user_study}
\vspace{-1mm}
\end{figure*}

\subsection{User Study}

To evaluate the generation quality of different talking head methods from a subjective human perspective, we conducted a user study involving 16 participants, as illustrated in Fig.~\ref{user_study}.
This study compares our method with five representative baselines, including Wav2Lip~\cite{prajwal2020lip}, AD-NeRF~\cite{guo2021ad}, RAD-NeRF~\cite{tang2022real}, TalkingGaussian~\cite{li2024talkinggaussian}, and GaussianTalker~\cite{cho2024gaussiantalker}. We trained models for four identities, and for each speaker, one video was synthesized using each method, resulting in 24 anonymized video clips. These clips were randomly shuffled and presented to participants in a blind test to eliminate bias.
In each trial, participants were presented with six video clips corresponding to the same identity, each generated by a different method, and were asked to score them comparatively.
Participants evaluated each clip along three perceptual criteria, \emph{i.e.}, \emph{Image Fidelity}, \emph{Video Realness}, and \emph{Audio-Visual Synchronization}, and assigned a rating from 1 to 6 for each method under each criterion (higher is better).
We aggregated the rating scores and visualized the average scores in a bar chart as shown in Fig.~\ref{user_study}. Results show that our method consistently outperforms other baselines across all evaluation dimensions.
This study also reveals that even some methods like Wav2Lip achieving high quantitative scores in Audio-Visual Synchronization, it may receive low user ratings on this metric due to their poor visual quality, highlighting the discrepancy between automated measurements and human perceptual evaluation.
Overall, the user study further validates the strength of our method in generating visually realistic and temporally coherent talking head videos.

\subsection{Ethical Consideration}
Talking head synthesis technology, while offering significant potential for advancements in digital communication, entertainment, and human-computer interaction, also introduces notable ethical concerns. 
These include the risk of malicious applications such as deepfake generation, the spread of misinformation, and unauthorized replication of personal identities—issues that stem from the technology’s capacity to generate highly realistic and photorealistic facial animations. Left unchecked, such misuse could erode public trust in digital media and contribute to broader societal harms.
We are committed to combating these malicious activities and promoting positive applications of the talking head synthesis technology.
We also actively support the development of robust forgery detection frameworks~\cite{prashnani2024generalizable,hua2023learning,xia2024inspector,shahreza2023comprehensive} to enhance security and authenticity in digital media.
In line with these principles, we will require authorization for code access, ensuring that it is only utilized for legitimate research or educational purposes. and watermark the generated videos to clearly distinguish synthetic content from the real one. Additionally, we suggest adding watermarks to all the generated videos to clearly delineate synthetic content from real-world footage. These measures are designed not only to protect against potential misuse but also to foster transparency and accountability in the development and application of talking head technologies.

\section{Conclusion}
In this paper, we have proposed a real-time talking head synthesis method based on a dynamic Gaussian Splatting framework. We develop the Audio Factorization Plane, which provides a compact and interpretable audio-driven dynamic scene representation. Additionally, an audio-guided saliency splatting method is designed to explicitly supervise the dynamism of 3D Gaussian points, thereby enhancing the modeling of dynamic facial regions while reducing
redundant complexity in less deformable areas. Our method can generate vivid and realistic talking videos with accurate audio-lip synchronization in real time, providing a strong baseline for high-quality and efficient audio-driven facial animation.


{\small
\bibliographystyle{IEEEtran}
\bibliography{IEEEabrv,main}
}

\vspace{-10mm}
\begin{IEEEbiography}
[{\includegraphics[width=1in,height=1.25in,clip,keepaspectratio]{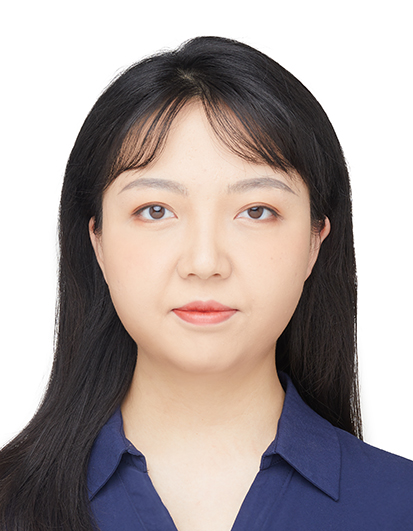}}]{Shuai Shen} received the B.S. and Ph.D. degree from the Department of Automation, Tsinghua University, China, in 2019 and 2024, respectively. She is currently working as a research fellow with the School of Electrical and Electronic Engineering (EEE), Nanyang Technological University, Singapore. Her research interests include visual generation, neural rendering, and face clustering. She serves as a regular reviewer member for TPAMI, TIP, TCSVT, TMM, CVPR, ICCV and ECCV.
\end{IEEEbiography}

\begin{IEEEbiography}
[{\includegraphics[width=1in,height=1.25in,clip,keepaspectratio]{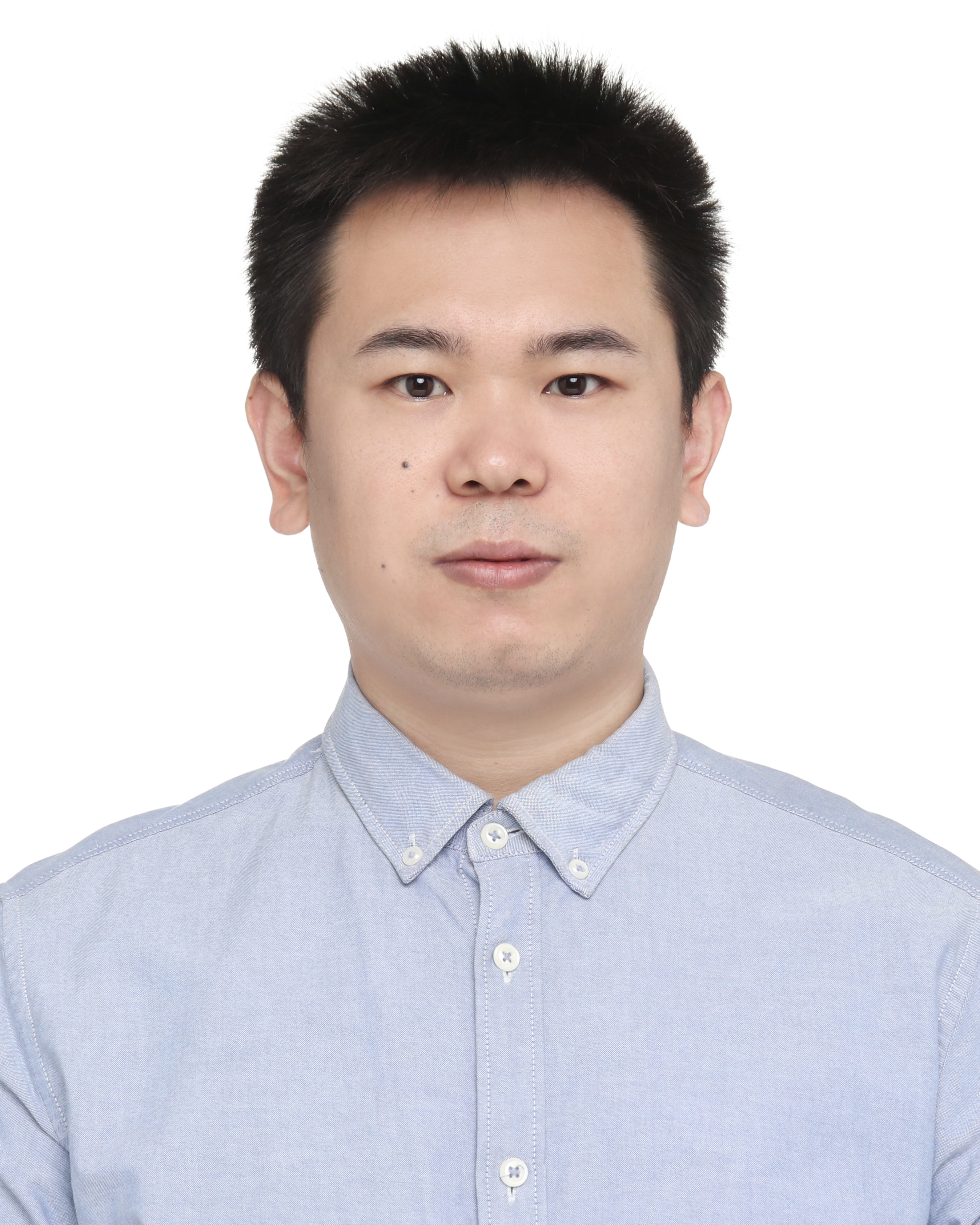}}] {Wanhua Li} (Member, IEEE) received the B.S. degree from the School of Data and Computer Science, Sun Yat-sen University, Guangzhou, China, in 2017, and the Ph.D. degree from the Department of Automation, Tsinghua University, China, in 2022. He is currently a Postdoctoral Fellow with the John A. Paulson School of Engineering and Applied Sciences at Harvard University, Cambridge, MA, USA. He has authored more than 20 top-tier conference/journal papers in CVPR, ICCV, ECCV, ICLR, NeurIPS, and TPAMI. His research interests include computer vision and machine learning, with particular expertise in neural rendering and vision-language models. He is as a regular reviewer member for TPAMI, TIP, TNNLS, TCSVT, TMM, IJCV, ICCV, CVPR, ECCV, and NeurIPS.
\end{IEEEbiography}

\begin{IEEEbiography}
[{\includegraphics[width=1in,height=1.25in,clip,keepaspectratio]{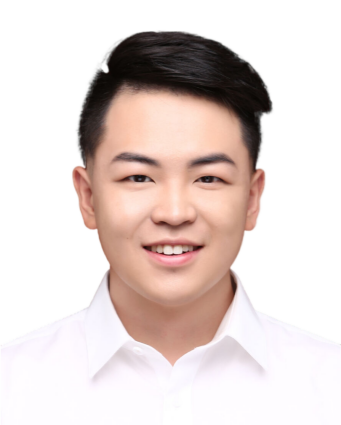}}]{Yunpeng Zhang} received the B.S. and M.Sc. degrees in Automation from Tsinghua University in 2019 and 2022, respectively. He is currently an algorithm engineer at PhiGent Robotics Co., LTD. His main research interests include monocular 3D object detection, multi-camera based 3D object detection, vision-based occupancy prediction, and end-to-end autonomous driving.
\end{IEEEbiography}

\begin{IEEEbiography}
[{\includegraphics[width=1in,height=1.25in,clip,keepaspectratio]{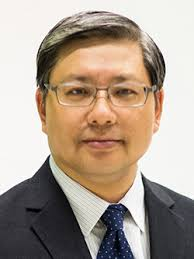}}]
{Yap-Peng Tan} (Fellow, IEEE) received the BS degree
from National Taiwan University, Taipei, Taiwan, in
1993, and the MA and PhD degrees from Princeton
University, Princeton, New Jersey, in 1995 and 1997,
respectively, all in electrical engineering. He is currently a full professor and the chair of the School
of Electrical and Electronic Engineering, Nanyang
Technological University, Singapore. His research interests include image and video processing, computer
vision, pattern recognition, and data analytics. He
served as an associate editor of IEEE Transactions on Circuits and Systems for Video Technology, IEEE Signal Processing Letters, and IEEE Transactions on Multimedis, as well as an editorial board member of the EURASIP Journal on Advances in Signal Processing and EURASIP Journal on Image and Video Processing. He was the Technical Program co-chair of the ICME 2015 and the ICIP 2019, and the general co-chair of the ICME 2010 and the VCIP 2015.
\end{IEEEbiography}

\begin{IEEEbiography}
[{\includegraphics[width=1in,height=1.25in,clip,keepaspectratio]{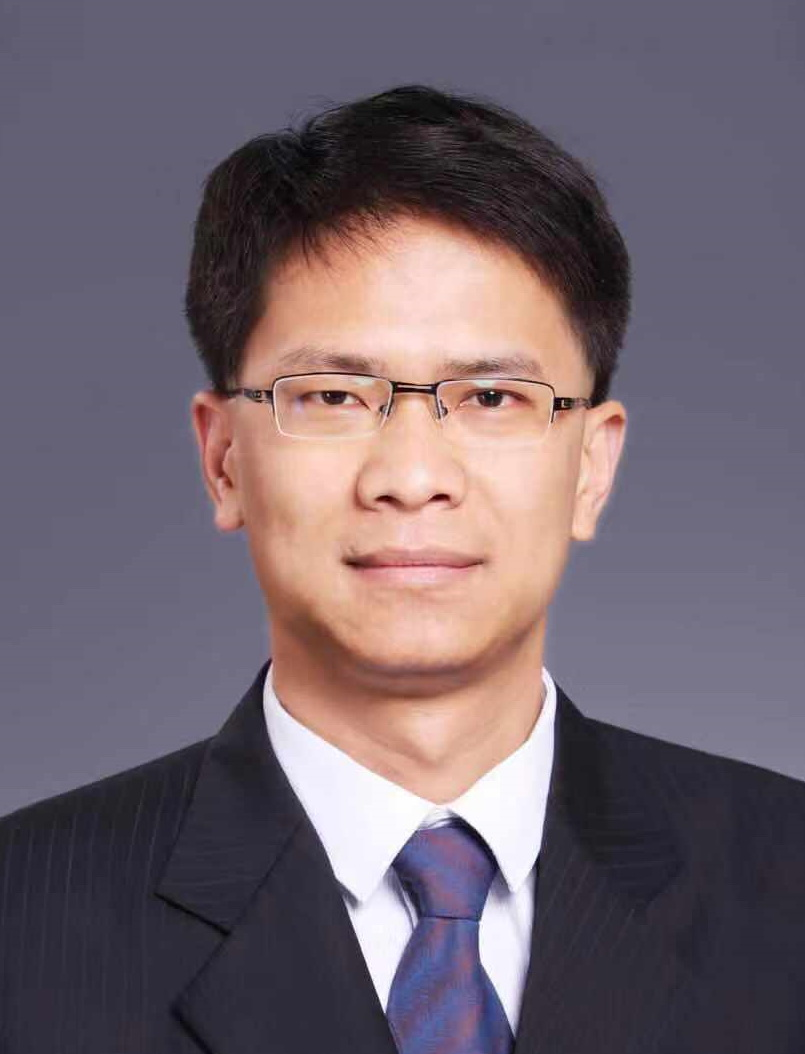}}]
{Jiwen Lu} (Fellow, IEEE) received the BEng degree
in mechanical engineering and the MEng degree in
electrical engineering from the Xi’an University of
Technology, Xi’an, China, in 2003 and 2006, respectively, and the PhD degree in electrical engineering from Nanyang Technological University, Singapore, in 2012. From 2011 to 2015, He was with the Advanced Digital Sciceces Center, Singapore. In November 2015, he joined the Department of Automation, Tsinghua University, where he is currently a
full professor and the deputy chair of the department.
His current research interests include computer vision, pattern recognition, multimedia computing, and intelligent robotics. He serves as the co-editor-of-chief for Pattern Recognition Letters, an associate editor for the IEEE Transactions on Image Processing, IEEE Transactions on Circuits and Systems for Video Technology, and IEEE Transactions on Biometrics, Behavior, and Identity Sciences, and Pattern Recognition. He was a recipient of the National Natural Science Funds for Distinguished Young Scholar. He is an
IEEE/IAPR Fellow.
\end{IEEEbiography}

\end{document}